# Photon-number-resolving single-photon detector with a system detection efficiency of 98% and photon-number resolution of 32

**Chaomeng Ding,**[a,b,c,#] **Xingyu Zhang,**[a,b,#] **Jiamin Xiong,**[a,b] **You Xiao,**[a,b] **Tianzhu Zhang,**[a] **Jia Huang,**[a,b] **Hongxin Xu,**[a,b,c] **Xiaoyu Liu,**[a,b] **Lixing You,**[a,b,c] **Zhen Wang,**[a,b] **Hao Li,**[a,b,d,*]

[a]*Shanghai Key Laboratory of Superconductor Integrated Circuit Technology, Shanghai Institute of Microsystem and Information Technology, Chinese Academy of Sciences, Shanghai 200050, China*
[b]*National Key Laboratory of Materials for Integrated Circuits, Shanghai Institute of Microsystem and Information Technology, Chinese Academy of Sciences, Shanghai 200050, China*
[c]*Center of Materials Science and Optoelectronics Engineering, University of Chinese Academy of Sciences, Beijing 100049, China*
[d]*Shanghai Research Center for Quantum sciences, Shanghai 201315, China*
[#]*These authors contributed equally*

**Abstract**. Efficiently distinguishing photon numbers is a crucial yet challenging technology for various quantum information and quantum metrology applications. While superconducting transition edge sensors offer good photon-number-resolving (PNR) capabilities, they are hampered by low detection speed, timing jitter, and complex cooling and readout requirements. In this work, we present a significant advancement toward achieving high-fidelity PNR single-photon detectors. The unique twin-layer configuration of superconducting nanowire atop a dielectric mirror ensures the near-unity detection efficiency. The segmented design enables spatial multiplexing, establishing a mapping relationship between pulse amplitude and registered photons. The fabricated detector exhibits impressive performance metrics, including a single-photon system detection efficiency (SDE) of ~ 98% at a dark count rate of 20 cps and photon-number resolution capability up to 32. Further characterization through detector tomography reveals high fidelities for two-, three-, and four-photon events, approximately 87%,73%, and 40% respectively. Moreover, the detector operates at a high count rate of 41 MHz at 3dB-SDE, with a low timing jitter of as low as 40 ps. With its near-unity efficiency, high photon-number resolution, low dark count rate and fast detection speed, we expect significant interest in these detectors, promising substantial benefits for weak light detection and optical quantum information applications.

**\*** Hao Li**,** E-mail: lihao@mail.sim.ac.cn

## 1 Introduction

Optical quantum states consist of superpositions of photon-number states. Achieving high-fidelity photon-number-resolving (PNR) single-photon detectors capable of efficiently discriminating and detecting these states has long been a key objective in photon detection, widely recognized for its critical role not only in fundamental research[1], but also for applications in quantum optics and quantum information, such as linear optics quantum computation[2,3], Gaussian boson sampling[4-6], quantum communication[7], and quantum metrology[8-10]. Thus far, superconducting transition edge sensors have emerged as a promising PNR detector capable of inherently resolving multiple photons, achieving high system efficiency above 90% at a wavelength of 1550 nm and a maximum resolution of up to 20[11,12]. However, limitations such as low count rates (typically below 1 MHz), modest timing resolution (~ ns), and the requirement for ultra-low operating temperatures (~100 mK) hinder their deployment in high-repetition-rate experiments and limit scalability. An alternative approach to PNR detectors involves using conventional single-photon detectors (SPDs) alongside spatial or temporal multiplexing techniques[13-18]. Here, multiple incident photons are distributed across a series of spatial or temporal modes and subsequently detected independently. High-fidelity detection of multiphoton events depends on achieving high single-photon system detection efficiency (SDE) and ensuring a sufficient number of distributed modes to minimize the chances of multiple photons occupying the same mode.

Semiconductor SPDs are commonly used in the development of PNR detectors but suffer from low detection efficiency owing to their limited material band gap[19]. Superconducting nanowires have emerged as promising candidates for detectors in quantum information systems[4,20] because of their high sensitivity and rapid response. The advancement of quantum information technology has driven the rapid development of superconducting nanowire SPDs (SSPDs) toward achieving near-unity system efficiency (≥98%)[21-23], low timing jitter (≤20 ps)[24-26], low dark count rates (≤100 Hz) , and high count rates (≥10 MHz) in the past two decades[27,28]. Meanwhile, while SSPDs have traditionally been considered 'binary' detectors capable of distinguishing only 0 and ≥1 photons, various approaches have been proposed to increase photon-number resolution. These include methods such as directly analyzing the responding pulses of conventional SSPD under multi-photon illumination[13-16], utilizing on-chip optical waveguide structures for temporal multiplexing schemes[17], and employing shunted resistors for spatial multiplexing schemes[17]. The signal-to-noise ratio of conventional SSPD devices can be enhanced through impedance-matching taper[14], low-timing-jitter devices[15], and large-inductance microstrips[16]. However, the reported maximum number of resolved photons does not exceed 10. The temporal multiplexing scheme based on waveguide structures can resolve up to 100 photons[17]. Yet, it faces challenges in achieving high-fidelity detection owing to optical coupling losses and limitations in detection speed arising from temporal distribution. In contrast, the parallel resistive PNR[29] based on spatial multiplexing technology[29,30] has shown high dynamic range PNR capabilities, simultaneously resolving up to 24 photons.

With advancements in efficiency and PNR approaches, the search for high-fidelity PNR detectors achieving near-unity efficiency and substantial photon-number resolution is ongoing. Herein, we make significant advances toward high-PNR-fidelity detectors. We employ a sandwiched twin-layer superconducting nanowire structure atop dielectric mirrors to ensure near-unity detection efficiency and high signal-to-noise ratio readout. A spatial multiplexing scheme utilizing multiple shunted resistors maps single-channel output pulse amplitudes to registered photon numbers. The fabricated detector exhibits a high single-photon SDE of 98% at a dark count rate of 20 cps and a photon-number resolution of up to 32. Detector tomography reveals detection fidelity rates of 87% for two-photon events and 73% for three-photon events. Additionally, the detector achieves a count rate of 41 MHz at 3dB-SDE and a timing jitter of 40 ps. Such detectors are expected to find applications in optical quantum computing and quantum metrology.

## 2   Results

*2.1 Device SDE and PNR resolution*

Figure 1(a) shows the detector architecture, featuring a segmented, sandwiched twin-layer superconducting nanowire positioned atop a distributed Bragg reflector (DBR) with a central reflection band at 1550 nm. The nanowire spans a sensing area approximately 20 μm × 20 μm in size, with a width/pitch of 80/160 nm, and is segmented into 32 pixels. These pixels are connected in series, and each is shunted by a neighboring resistor of 40 Ω. The adoption of the sandwiched twin-layer superconducting nanowire structure aims to enhance the optical absorption properties of the nanowires[21] and enhance the signal-to-noise ratio (SNR) of photon responses to achieve better photon-number resolution. The segmented pixels and neighboring resistors constitute a spatial multiplexing scheme that maps output pulse amplitudes to registered photons. When the incident photons trigger certain pixels of the detector, the resistance generated by hot spots exceeds that of the shunt resistor, allowing bias current to flow through the shunt resistor to produce an output signal. Owing to the proximity of shunt resistances across different pixels, the output signal amplitude is predominantly linearly correlated with the number of incident photons. One advantage of this type of multiplexing is its compatibility with conventional device designs and fabrication processes, preserving detector efficiency. Additional details and an image of the device and nanowire can be found in Supplementary Note S1.

Figure 1(b) illustrates the curve of single-photon SDE and dark count rate of the device, characterized under a setup similar to that described in a previous study[21]. The detector exhibits a broad saturation range, maintaining an average efficiency of 97.5% at 1555 nm wavelength, with a minimum of 97.0% and a maximum of 98.6%, alongside a dark count rate of 20 cps. At 1550 nm, the detector achieves a single-photon SDE of 93.7%. The total uncertainty of the detector is ~0.65%, which consists of the uncertainty of the power meter, the overall stability of the optical path (including the light source, attenuator and polarization controller), the uncertainty of photon response count, the uncertainty of the fiber welding loss and the uncertainty of the polarization. The detailed device calibration and measurement setup can be found in Supplementary Note S2.

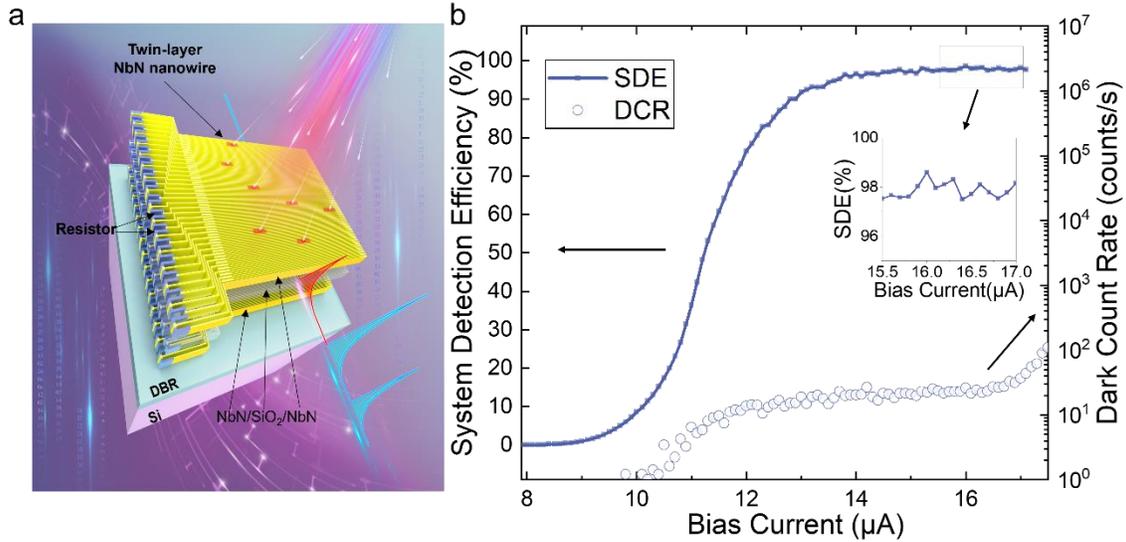

**Fig.1** Detector architecture and efficiency performance. (a) Diagram of detector structure. The NbN/SiO$_2$/NbN nanowire structure and shunted resistors are deployed on a distributed Bragg reflector fabricated on a silicon substrate. The nanowire covers a sensing area of approximately 20 μm × 20 μm with a width/pitch of 80/160 nm. It is divided into 32 pixels connected in series, each shunted by a neighboring resistor of 40 Ω. Additional details and an image of the device and nanowire can be found in Supplementary Note S1. (b) System detection efficiency and dark count rate vs. bias current, measured at 0.85 K. The detector demonstrates a broad saturation range, achieving an average efficiency of 97.5% at a wavelength of 1555 nm and a dark count rate of 20 cps.

To evaluate the PNR capability of the device, we first conducted SPICE model simulations[31], illustrating a linear increase in pulse amplitude with registered photons, as shown in Fig. 2(a). We construct 32 nanowires and connect each of them in parallel to a resistor. A bias current is applied to the detector, and a current source is used to simulate the optical input to make the nanowire lose its superconducting state. The simulation includes a 58 dB gain from the amplifier and rectifier circuit. The designed amplitude increment (~16 mV) per photon is greater than the system noise to distinguish between different photons, and it can be adjusted by varying the bias current, parallel resistance and amplification circuit. Subsequently, experimental characterization of PNR capability followed the setup depicted in Fig. 2(b). The arbitrary function generator (Tektronix, AFG3252C) provides a 100 kHz pulse signal to modulate the picosecond pulse light source, and the optical attenuator (EXFO, FVA-3150B) regulates the total number of incident photons. The average number of photons per pulse μ is equal to the ratio of the total number of incident photons to the pulse frequency. A polarization controller optimized light polarization for maximum detection efficiency. By varying the average photon number per pulse via attenuators, up to 32 distinct voltage amplitude waveforms were observed on an oscilloscope, as illustrated in Fig. 2(c), demonstrating a photon-number resolution capability of 32.

Figure 2(d) shows the histogram of output pulse amplitudes at an incident mean photon number of μ = 5, revealing 8 distinct Gaussian peaks. By scanning the light intensity, 32 Gaussian peaks can be seen. Additional results on histograms for different μ values can be found in Supplementary Note S3. To assign a detected event with a Gaussian peak to a specific photon number n, we define voltage blocks such that if $V_n \leq V \leq V_{n+1}$, we assign the event to photon number n[32]. To assess the accuracy of this voltage-based photon-number assignment, we calculate the proportion of the area under each Gaussian peak in its assigned voltage block. The device's high SNR facilitates a high assignment probability, as depicted in Fig. 2(e). For n $\leq$ 6, the probability of accurate assignment exceeds 0.99. This near-perfect acquisition of output information is enabled by well-separated output distributions.

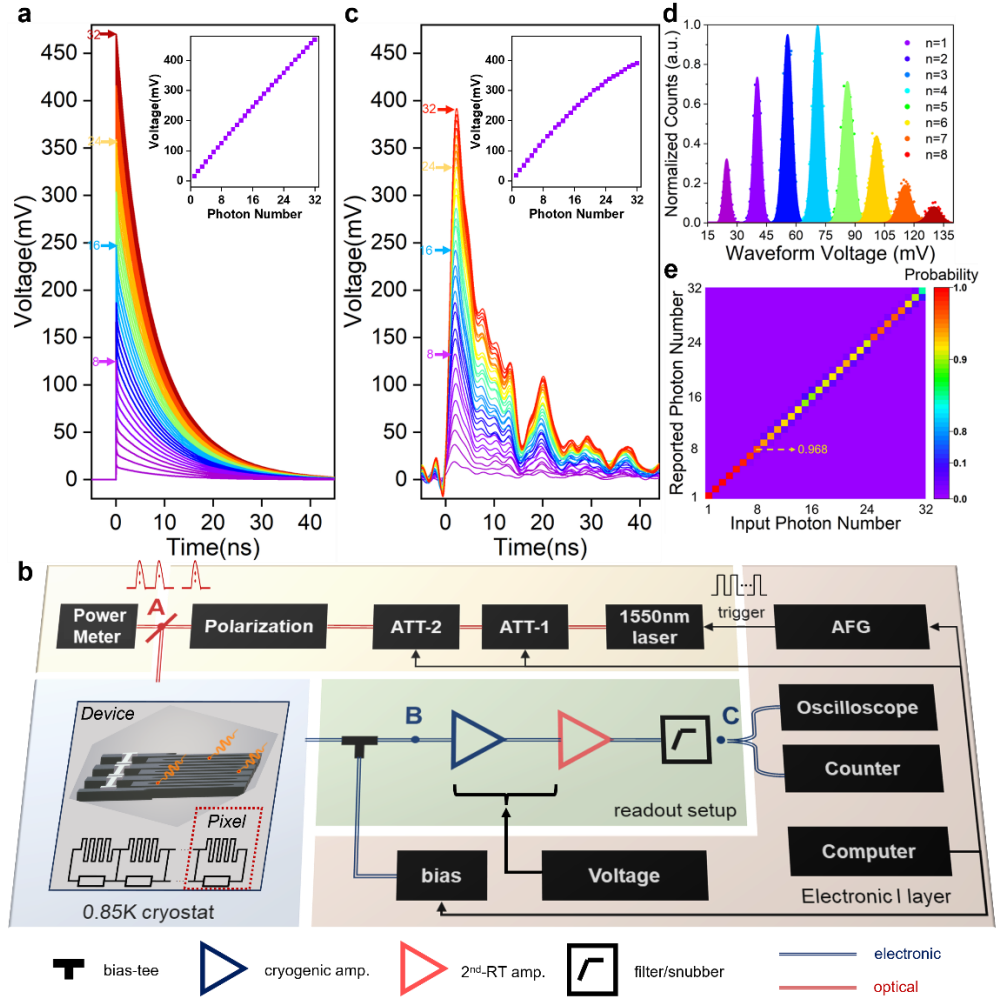

**Fig.2** Readout capability characterization. (a) Waveform simulation diagram. SPICE model simulates output waveforms for 1–32 photons. The illustration shows the relationship between photon number and amplitude. (b) Experimental setup. An arbitrary function generator (AFG) modulates a pulsed light source, with intensity controlled by two attenuators (ATT). After polarization controller adjustment and optical power calibration, input light A is incident on the detector. The output signal B generated by the detector response flows through the cryogenic amplifier, the secondary amplifier, and the high-pass filter to form the final output C. (c) Device waveform diagram. The waveforms from bottom to top were measured by an oscilloscope at a bias current of 16.0 µA, corresponding to 1-32 photon events. The illustration shows the relationship between photon number and amplitude. (d) Output amplitude distribution. Device measured at µ = 5 shows eight Gaussian peaks corresponding to 1–8 photons. (e) Gaussian peak distribution of amplitude corresponding to different photons. The panel highlights cases with the most overlap of Gaussian peaks. For n = 1–6, the probability of accurately resolving Gaussian peaks exceeds 99%.

## 2.2 Detector tomography and POVM reconstruction

The detector tomography involves reconstructing the positive operator-valued measure (POVM) elements[33,34], which indicate the probability of the device outputting n photons given m incident photons. Experimentally, multiple sets of input pulsed optical signals at a frequency of 100 kHz, parameterized by the average photon number µ and following different Poisson distributions, are organized into an input matrix I. We select µ values ranging from 0.1 to 5 to ensure that light intensities for n = 0–6 cover most scenarios, minimizing measurement errors. The output matrix O records counts of pulse amplitudes using a pulse counter across different µ values. Note that there is a comparator inside the counter and it generates a count when the voltage signal is higher than the comparison voltage. Figure 3 illustrates the measured counts as a function of comparison level at µ = 1, with the inset displaying the output photon-number distribution extracted from the curve, where the count between the two adjacent counting platforms is the total number of n-photon events. Additional measurement results spanning µ values from 0.1 to 5 are provided in Supplementary Note S4.

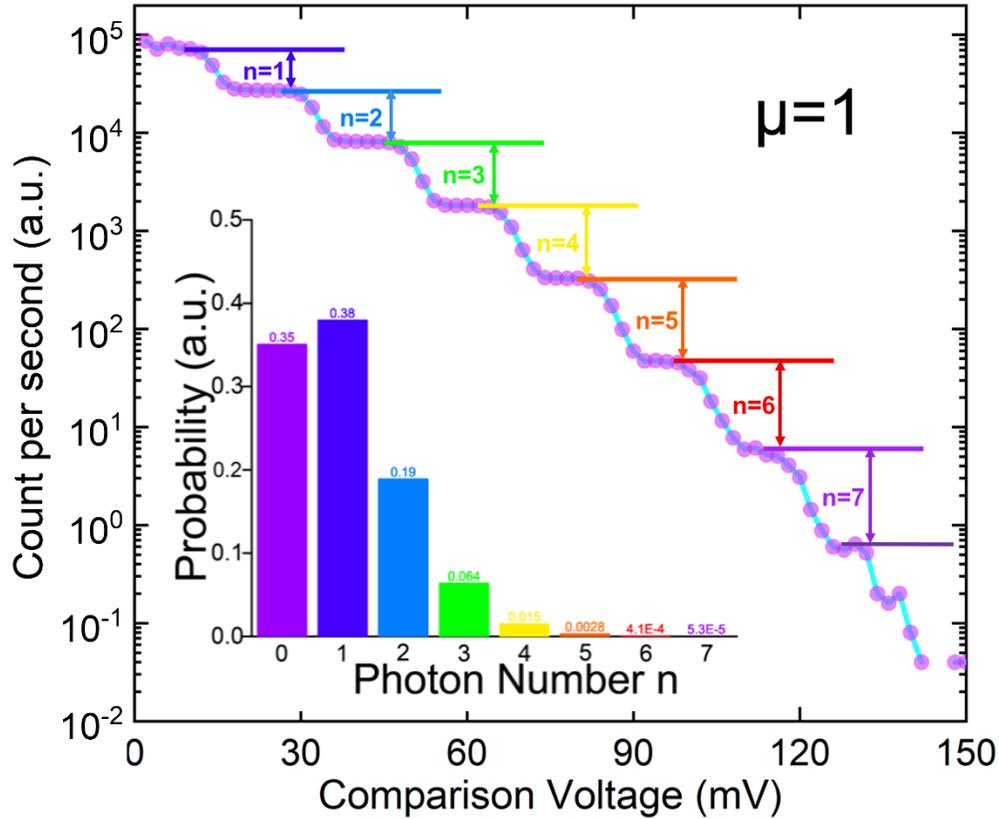

**Fig.3** Output acquisition and processing. The counts and comparison level curve is shown at a bias current of 15.0 µA, with an average photon number per pulse of µ = 1. This curve enables the extraction of the output photon-number distribution which represents the photon-number probabiltiy in the laser pulse as given by the inset.

Given the matrix of input states I and outcomes O, the fidelity matrix P corresponding to the POVM set is reconstructed by finding the minimum of $||P * I - O||_2$ [35]. The POVM element $P_{nm}$ represents the probability of registering an n-click when m photons are incident. While the m can in principle take an infinite value, in practice it can be truncated to finite values[36]. Here we truncate m to the value 12, and the mean photon number of µ=0.1-5 is enough to cover the incident photon-number events. To ensure the POVM is physically realisable, the constraints are used during the reconstruction, where all the POVM elements are non-negative, the elements in the lower triangle are 0, and the sum of the elements in each column equals to 1. Table 1 shows the fidelity matrix for six photon-number events, with diagonal elements indicating probabilities of accurate detection (the full fidelity matrix is given in Supplementary Note S5). The columns represent the input photon number m, and the rows represent the output photon number n. The device achieves a single-photon detection fidelity of 97.5%, a two-, three- and four-photon fidelities of 87.4%, 73.4%, and 40.5%. These multiphoton detection fidelity performances significantly surpass previously published results. A comparison is given in Supplementary Note S6.

**Table 1. Fidelity matrix P.**

| P | | Input photons m | | | | | | |
|---|---|---|---|---|---|---|---|---|
| | | 0 | 1 | 2 | 3 | 4 | 5 | 6 |
| Output photons n | 0 | 1.000 | 0.025 | 0.043 | 0.053 | 0.074 | 0.085 | 0.090 |
| | 1 | 0 | 0.975 | 0.082 | 0.079 | 0.125 | 0.132 | 0.118 |
| | 2 | 0 | 0 | 0.874 | 0.135 | 0.191 | 0.195 | 0.167 |
| | 3 | 0 | 0 | 0 | 0.734 | 0.205 | 0.132 | 0.111 |
| | 4 | 0 | 0 | 0 | 0 | 0.405 | 0.186 | 0.136 |
| | 5 | 0 | 0 | 0 | 0 | 0 | 0.271 | 0.164 |
| | 6 | 0 | 0 | 0 | 0 | 0 | 0 | 0.215 |

To verify the accuracy of the fidelity matrix, quantum state reconstruction using the fidelity matrix is performed, comparing the expected input distribution with the reconstructed distribution. The Hellinger distance is employed to quantify the similarity between the two probability distributions[37]. This metric, commonly used for comparing probability distributions, describes the difference between the reconstructed input $P_R = (P_{R1}, ... P_{Ri})$ and the expected input $P_E = (P_{E1}, ... P_{Ei})$ as $1 - H^2 = \sum \sqrt{P_{Ri} P_{Ei}}$. An H value closer to 0 indicates greater similarity between the distributions. For µ values less than 1.5, the reconstructed result closely matches the actual input, with H values typically less than ~0.1. More details are given in Supplementary Note S5.

## 2.3 Timing jitter and counting rates

Figure 4(a) illustrates the single- and multiphoton time jitter of the detector. As the output amplitudes vary with input photons, the detector demonstrates reduced timing jitter with increasing photon number. Specifically, the jitter is 382 ps for photons n = 1, 234 ps for n = 2, and 40.4 ps for n = 32. In this experiment, an fs-pulsed laser (FPL-01CAF, Calmar) coupled with an electro–optic modulator (FTM7937EZ/202) is a low timing-jitter pulsed laser source. We adjust light intensity and configure the oscilloscope to capture waveforms in the desired amplitude range. Figure 4(b) shows the count rate curve of the detector, reaching approximately 41 MHz when efficiency decreases by 3 dB. Notably, the measurement includes continuous laser light and a bias current of 15.0 µA applied to the detector.

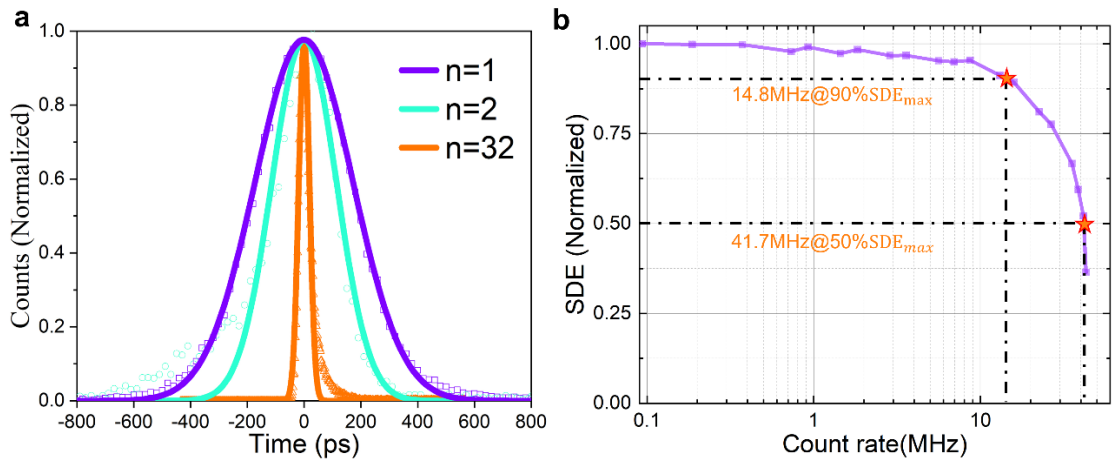

**Fig.4.** Time jitter and count rate. (a) Timing jitter for different photons. FWHM is $n = 1$ (Violet), 382 ps; $n = 2$ (Cyan), 234 ps; $n = 32$ (Orange), 40.4 ps. (b) Count rate and efficiency curve. The count rate reaches 41 MHz, with the SDE decreasing to −3 dB.

## 3   Discussion

### 3.1   Photon number resolution

Our detector uses pulse amplitude to map incident photon numbers, making SNR crucial for achieving high photon-number resolution. While our detector currently achieves a photon-number resolution of 32, further enhancement is anticipated. The following methods are feasible. 1) Enhancing the SNR by increasing the switching current of the nanowire. Our detector operates with a switching current of ~18 µA, which can be optimized through enhancements in the device substrate, superconducting material, and nanowire fabrication techniques. A 25–30 µA nanowire switching current can potentially increase the photon-number resolution to >50 at 1550 nm. 2) Combining spatial and temporal multiplexing methods. Time-multiplexed readout has been employed for photon-number resolution, enabling the detection of >100 resolved photons[17], albeit at the cost of reduced count rates. 3) Utilizing photon information from output pulses, including but not limited to the amplitude. Conventional SSPDs distinguish between different photon incident numbers based on the rising edge of the output signal. Leveraging this edge information could double or triple the maximum number of resolvable photons. Challenges remain, such as optimizing the readout of the rising edge and handling acquired data, but resolving these issues is expected to significantly improve SSPD resolution.

*3.2 Multiphoton detection fidelity*

The PNR detection fidelity, defined as the probability of accurately reconstructing the incident photon-number state, is used to evaluate the performance of detectors. As given above, the probability of detecting *n* photons when *m* photons are incident is denoted by $P_{nm}$. The detection fidelity for *n* photons in a ideal multipixel photon-number resolution device is expressed as $P_{nn} = (\frac{\eta}{N})^n \frac{N!}{(N-n)!}$ for *n* < *N*, where *N* represents the total number of pixels of the device, and $\eta$ represents the efficiency of the device[38]. High SDE and numerous spatial pixels are essential for achieving high-fidelity detection of multiphoton events. Greater efficiency and pixel count correlate directly with improved detection fidelity, the detailed relationship is described in Supplementary Note S7. Using spatial multiplexing, the current detector achieves significant photon-number resolution, up to 32. It maintains a near-unity SDE of 97.5% via a segmented, sandwiched superconducting nanowire structure. This setup enables detection fidelity of 87.4%, 73.4%, and 40.5% for two-photon, three-photon, and four-photon events, respectively, as detailed in Table 1. The high detection fidelity in detecting multiphoton events and a high count rate support emerging quantum information applications such as Gaussian boson sampling and quantum metrology[39]. Further enhancing multiphoton detection fidelity is desirable through increased photon-number resolution and optimized SDE, as described above.

It is worth mentioning that the detection fidelity observed through detector tomography tends to be lower than that predicted by the formula provided above. This discrepancy arises primarily because incident light follows a Gaussian distribution, which is not uniformly distributed across the detector pixels. The smaller pixel size and the increased number of pixels are required to further improve the multiphoton event detection fidelity.

*3.3 Detector crosstalk*

The potential crosstalk resulting from heat transfer or current distribution between pixels is an important consideration that can impact the distribution of photon numbers in detection results. We define single-pixel crosstalk probability as $p_{xtalk} = p(2|1) - p(1|1)p(1|0)$, where p (2|1) is the probability that one incident photon triggers a two-photon click, p (1|1) denotes one-photon detection fidelity, and p (1|0) represents the dark count probability when no photons are incident[35]. In our experiments, we maintained an average photon number per pulse of μ = 0.01 to minimize the likelihood of multiphoton events. We measured the ratio of two-photon to single-photon counts across pulse frequencies ranging from 10 kHz to 100 MHz. By subtracting the inherent two-photon ratio in laser pulses, we found that device-induced crosstalk was less than 0.1%, rendering it negligible for our experiment. More details about the crosstalk are given in Supplementary Note S8.

In conclusion, we have demonstrated a single-photon PNR detector leveraging superconducting nanowire and spatial multiplexing technologies. This detector achieves nearly perfect SDE, large photon-number resolution, low dark count rates, and high-count rates. The enhanced photon-number resolution enables fundamental exploration of mesoscopic and macroscopic quantum effects, while the high-fidelity multiphoton detection promises advancements in emerging quantum information applications.

## 4 Materials and methods

### 4.1 Device fabrication

First, a NbN/SiO2/NbN sandwiched superconducting film, with layer thicknesses of 6/3/6 nm, was prepared on a Si substrate coated with a 13-periodic SiO2/Ta2O5 DBR structure. The second step involved defining the NbN pattern. First, a MARK pattern necessary for subsequent alignment was produced using laser direct writing. Subsequently, a deposition process employing a high vacuum evaporation system (ULVAC, EI 5) applied 8-nm-thick Ti followed by 50-nm-thick Au, which was then selectively removed. Next, using electron beam lithography, the nanowire area—covering a sensitive 20 μm × 20 μm square with a width/pitch of 80/160 nm—was exposed alongside auxiliary connection lines. Reactive ion etching (RIE) was utilized to etch out the pattern. A resistance pattern was produced using laser direct writing, followed by a second deposition and removal process involving 90-nm-thick Ti. The Ti layer exhibited a sheet resistance of approximately 26 Ω/sq. Finally, electrodes were defined using MA6 ultraviolet exposure and etched using RIE. Notably, exposure to air produced a thin oxidation layer of approximately 1 nm on the NbN surface. Before Ti resistor deposition using EI 5, Ar ion bombardment was employed to remove this oxide layer because it significantly increases the resistance between the Ti film and the NbN.

### 4.2 Device packaging

The device is affixed to a copper holder using low-temperature glue. One of the electrode is connected to the ground with bonding wires, and the other is connected to the SMA connector. Subsequently, the light spot is positioned precisely at the center of the photosensitive area, and the packaged device is cooled to 0.85 ± 0.05 K using an adsorption chiller. The detector's bias and readout are facilitated through a bias tee and coaxial cables with a 50 Ω impedance. The bias circuit features an isolated voltage source and a series resistor of 100 kΩ, establishing a quasiconstant current bias via the DC arm of the bias tee. To increase the SNR of the readout signal, a homemade amplifier was used to amplify the signal at the 40 K stage. The signal was then sent to room temperature for a second stage of amplification before being connected to the counter or oscilloscope.

*4.3 SDE measurement*

The detailed SDE measurement can be found in a published study[21]. The measurement settings are as follows: A continuous laser (Keysight, N7776C) is used as the input light of the detector, followed by two optical attenuators (EXFO, FVA-3150B) to control the incident photon number A polarization controller (Thorlabs, FPC561) is employed to adjust the polarization state to get the maximum efficiency. The fiber fusion splicing was used to avoid the optical loss caused by the conventional fiber connector. The input fiber was connected to the power-meter fiber by fusion splicing. After the calibration of the incident light power, we broke the input fiber and fusion spliced it to the SNSPD fiber. The SDE curve is obtained by changing the bias current. At each bias current, an automated shutter in a variable attenuator blocked the laser light, and the dark counts are collected for 10 s. The light is then unblocked, and the pulse counts are measured by collecting pulse counts for another 10 s. The SDE of the detector is defined as SDE = (PCR − DCR)/PR, where PCR is the output pulse count rate of the SNSPD measured using a pulse counter, DCR is the dark count rate, and PR is the photon rate input to the detection system.

**Data availability**

Data underlying the results presented in this paper are not publicly available at this time but may be obtained from the authors upon reasonable request.

**Funding**

We thank the National Natural Science Foundation of China (U24A20320, 62401554), Innovation Program for Quantum Science and Technology (No. 2023ZD0300100), and Shanghai Municipal Science and Technology Major Project (2019SHZDZX01) for their financial support.

**Disclosures**

The authors declare no conflicts of interest.

## Acknowledgments

H.L. conceived the idea of the detector. C.D. and X.Z. performed the numerical simulations, fabricated the device and implemented the measurement experiments. J.X., Y.X., T.Z., J.H., H.X. and X.L. helped with device fabrication and measurement. C.D., X.Z., and H.L discussed the results and wrote the manuscript with input from all authors. All the authors reviewed the manuscript. H.L. and L.Y. supervised the work. We also acknowledge the support from the Superconducting Electronics Facility (SELF) in SIMIT for device fabrication.

## Supplemental document

See Supplement 1 for supporting content.